\documentclass[preprint,aps,prl,amsmath,amssymb,showpacs,nofootinbib]{revtex4-1}

\usepackage{hyperref}

\usepackage{graphicx}	% for eps images
\usepackage{eucal}		% better fonts for \mathcal
\usepackage{mycommands}
\usepackage{etoolbox}  % for a LaTeX if statement

\newtoggle{geocoords}
\togglefalse{geocoords}  % use plasma coordinate conventions

\iftoggle{geocoords}{
	\newcommand{\LD}{L_d^{-2}}
	}{
	\newcommand{\LD}{\r_s^{-2}}
	}
\newcommand{\nablabarsq}{\ol{\nabla}^2}
\newcommand{\ybar}{{\ol{y}}}
\newcommand{\xbar}{{\ol{x}}}
\newcommand{\kbsq}{\ol{k}^2}
\newcommand{\N}{\mathcal{N}}

\providecommand{\eqnref}{}
\renewcommand{\eqnref}[1]{Eq.~\eqref{#1}}					% reference equation number easily referring only to the label
\newcommand{\figref}[1]{Figure~\ref{#1}}	% can change to "Fig." or "Figure" as desired

\begin{document}

%%% Title, Name %%%
\title{Dynamics of Zonal Flows: Failure of Wave-Kinetic Theory, and New Geometrical Optics Approximations}

\author{Jeffrey B.~Parker}
\email{parker68@llnl.gov}
\affiliation{Stanford Law School, Stanford, CA 94305}
%\thanks{ahhh} % goes at end of paper
%\affiliation{Lawrence Livermore National Laboratory, Livermore, CA 94550}

\date{\today}

\begin{abstract}
The self-organization of turbulence into regular zonal flows can be fruitfully investigated with quasilinear methods and statistical descriptions.  A wave kinetic equation that assumes asymptotically large-scale zonal flows is pathological. From an exact description of quasilinear dynamics emerges two better geometrical optics approximations. These involve not only the mean flow shear but also the second and third derivative of the mean flow. One approximation takes the form of a new wave kinetic equation, but is only valid when the zonal flow is quasi-static and wave action is conserved.
\end{abstract}

%  --- The abstract I would want to write, if I had more space
%The self-organization of turbulence into regular zonal flows can be fruitfully investigated with quasilinear methods.  A wave kinetic equation that assumes asymptotically large zonal flows is pathological.  From an exact description of quasilinear dynamics, we describe two simpler geometrical optics approximations that perform much better.  They involve not only the mean flow shear but also the second and third derivative of the mean flow.  The first formulation predicts the dispersion relation of zonostrophic instability remarkably well even when spatial scale separation is not satisfied.  The second takes the form of a new wave kinetic equation in terms of conservation of wave action, but is only valid when the zonal flow varies slowly.  In particular, it is not valid in the transient regimes in which the zonal flow grows and saturates because wave action is not conserved.

\maketitle

%%%% Body %%%%
% Put citations in PRL that are at the end of a sentence before the period \cite{styleguide}.
\paragraph{Introduction}
Self-organization and emergent phenomena in systems with many degrees of freedom and that are far from thermal equilibrium remain a frontier in physics.  In fluid dynamics, where turbulence and strong nonlinearities are ubiquitous, the difficulty is amplified because many standard techniques are unavailable.

A striking example of self-organization is zonal flow, which refers to banded flows alternating in space and quasistationary in time.  Zonal flow is formed from and coexists with turbulence in the diverse physical contexts of magnetically confined toroidal plasmas \cite{fujisawa:2009,hillesheim:2016}, planetary atmospheres \cite{vasavada:2005}, and possibly astrophysical discs \cite{johansen:2009, kunz:2013}.  Common to each of these physical systems are directions of symmetry, driven turbulent flow, and a gradient in the rotation, density, pressure, or other background quantity.

%  (Bai Stone 2014 -- stone calls them zonal flow but refers to density radial density variations.  Why does he use the term "flow" and call it a density variation?  I don't understand this -- is there an associated flow?) ---- ask Kunz if there are other higher quality articles I should cite.  Also, is there a wave associated with the Coriolis force in accretion disks?  A Rossby-wave equivalent?)

In recent years, a quasilinear approximation has facilitated progress in the fundamental understanding of zonal flows  \cite{farrell:2003,farrell:2007,marston:2008,srinivasan:2012,tobias:2013,parker:2013,parker:2014,constantinou:2014,bakas:2013b,bakas:2015,constantinou:2016}.  The quasilinear approximation involves neglecting the fluctuation self-nonlinearity while retaining the basic nonlinear coupling between mean flows and fluctuations.  This truncation eliminates the Kolmogorov cascades and wave-wave interactions, but numerical simulations of quasilinear dynamics show that \emph{fluctuations} can still drive the spontaneous formation of zonal flows \cite{srinivasan:2012}.  Hence, even though many fluid systems are naturally turbulent, turbulence per se is not a critical factor in driving zonal flows, and many aspects of the problem can be qualitatively understood in the simpler quasilinear setting.  This approach has found use beyond zonal flows for understanding generation of other large-scale structures such as the magnetorotational dynamo \cite{squire:2015} and rolls and streaks in wall-bounded shear flow \cite{farrell:2012}.  In these quasilinear models, fluctuations are typically driven by external white noise forcing, an analytically convenient assumption.

Because the quasilinear system couples the fluctuations with the mean field, the equations are \emph{stochastically linear}, despite being dynamically nonlinear.  One can therefore apply an averaging procedure and derive an equation of motion for the covariance without facing the standard closure problem of turbulence in which unknown triple correlations appear.

This averaging leads to a coupled set of equations for the two-point, one-time covariance of the fluctuations and the mean flow \cite{farrell:2003,farrell:2007,marston:2008,srinivasan:2012}.  These equations are called CE2, short for second-order cumulant expansion.  CE2 offers a set of nonlinear deterministic equations in which rapid fluctuations have been averaged away yet is still equivalent to the quasilinear dynamics, and is the simplest consistent statistical formulation in which to study inhomogeneous flows.

It is impossible to overstate the importance of CE2 in the context of the quasilinear model.  If statistical inhomogeneity exists in only one dimension, then in a particular limit and with a particular averaging procedure, CE2 describes the exact dynamics of a \emph{single realization}, not merely the statistics of an ensemble of trajectories \cite{srinivasan:2012,parker:2013}.  Some questions about the interpretation and relation of statistical ensembles to the actual dynamics of interest can therefore be avoided.  This limit is obtained when the domain size in the zonal direction is sufficiently large so that a zonal average becomes equivalent to an ensemble average over the noise, a particular form of ergodicity akin to a thermodynamic limit.  No assumption of separation of time or spatial scales is necessary.

% (Maybe) When the domain is finite, statistical fluctuations about the mean arise in the Reynolds stresses, and CE2 no longer is exact.  Strong nonergodicity can cause zonal jets to bounce between metastable states or even prevent coherency.

With CE2, a basic theoretical framework of zonal flows has been uncovered \cite{farrell:2003,farrell:2007,srinivasan:2012,parker:2013, parker:2014}.  A statistically homogeneous equilibrium consisting of fluctuations without zonal flows always exists, though it may be unstable.  If the incoherent fluctuations are sufficiently intense, they can drive a symmetry-breaking instability that grows into zonal flow.  This instability is known as the \emph{zonostrophic instability} and provides an emergent length and time scale for zonal flows.  The instability also has a real eigenvalue, i.e., a perturbation is stationary as it grows.

The zonostrophic instability has been shown to be a generalization of what is sometimes called a modulational or secondary instability of a primary eigenmode \cite{gill:1974,lorenz:1972,connaughton:2010}.  When the background spectrum within zonostrophic instability is specialized to correspond with the primary mode, the dispersion relation agrees identically with previous results \cite{parker:thesis,parker:book,bakas:2015}.

When zonostrophic instability is marginally unstable, dynamics can be reduced to the real Ginzburg--Landau equation with universal behavior \cite{parker:2013, parker:2014}.  Qualitative features of the Ginzburg--Landau equation provide insight into behavior often observed in numerical simulations.  For example, merging zonal jets are commonly seen in the transient phase prior to saturation \cite{huang:1998,scott:2007}.  This phenomenon exists within the Ginzburg--Landau equation and can be related to jet stability.  As another example, some have remarked about the existence of multiple attractors or dependence on initial conditions \cite{danilov:2004,marcus:2012}.  This property, too, follows directly from the Ginzburg--Landau equation.  More generally, the pattern formation conceptual framework has proven useful \cite{cross:1993,cross:2009}.

All of these results have been understood within the quasilinear approximation using CE2 in the Charney--Hasegawa--Mima model.  While it has not yet been concretely demonstrated that the same structure exists in the original, fully turbulent system, some numerical evidence indicates it does \cite{parker:2014}.

\begin{figure}
		\centering
		\includegraphics[width=3.3in]{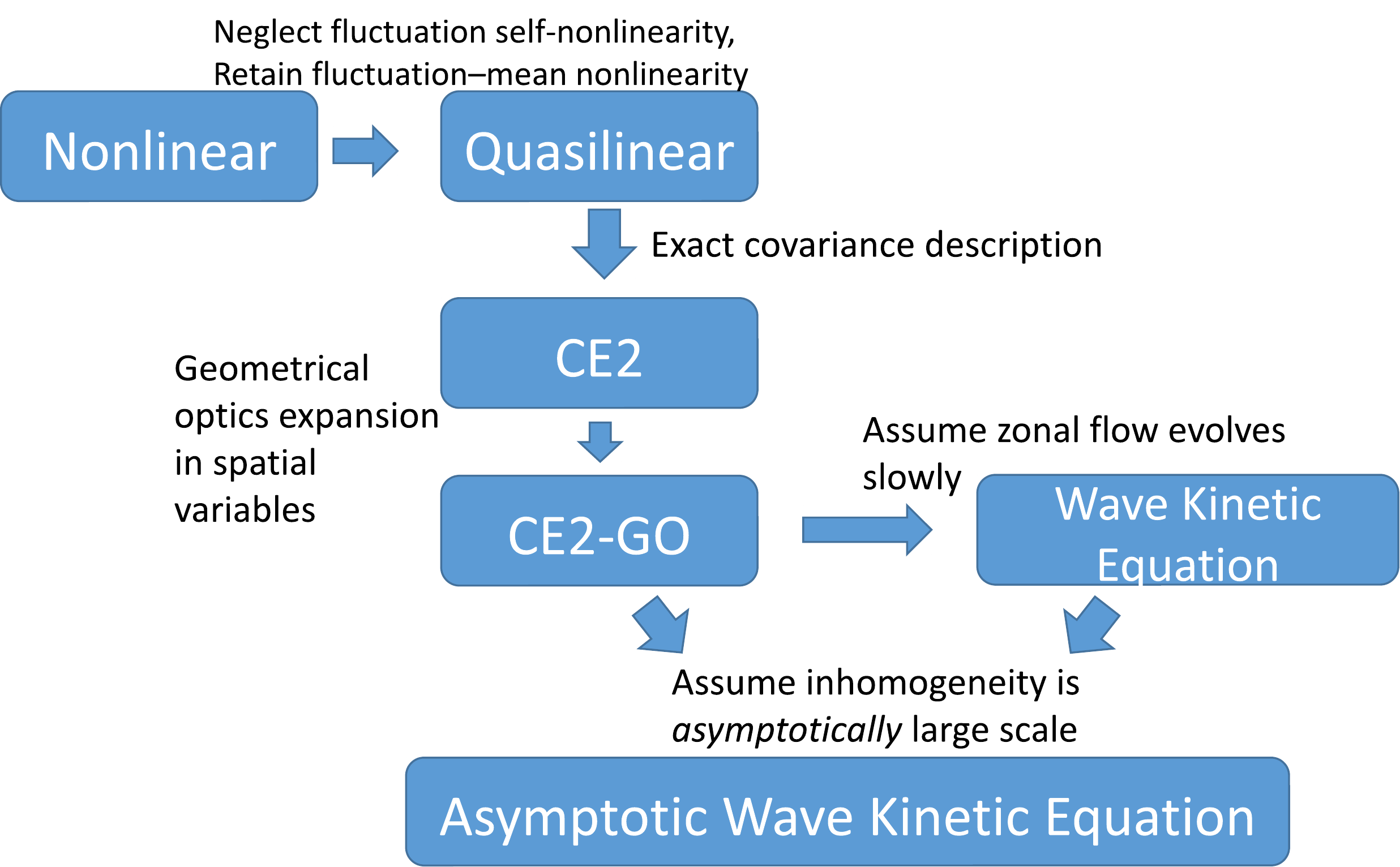}
		\caption{Hierarchy of models.}
		\label{fig:hierarchy}
	\end{figure}

In this paper, we discuss a few approximations to CE2 that offer the promise of a simpler, more intuitive form.  Two of these invoke wave kinetic equations, and we explain why this approach breaks down when studying zonal flow dynamics, although it may still prove useful in steady state.  We introduce another geometrical optics approximation that may be both accurate and intuitively appealing.  The relationship between these models is shown in \figref{fig:hierarchy}.

\paragraph{Quasilinear and CE2 equations of motion}
The 2D Charney--Hasegawa--Mima equation has been used to model turbulent flow in 
	\iftoggle{geocoords}{atmospheric fluid on a rotating planet (uniformly magnetized plasma with a density gradient)}{uniformly magnetized plasma with a density gradient (atmospheric fluid on a rotating planet)}.  We use the Generalized Hasegawa--Mima equation \cite{smolyakov:2000, krommes:2000},
	\begin{equation}
	\iftoggle{geocoords}{
		\partial_t \z(x, y, t) + \v{v} \cdot \nabla \z + \b \partial_x \psi = f + D,
	}{
		\partial_t \z(x, y, t) + \v{v} \cdot \nabla \z - \k \partial_y \psi = f + D,
	}
	\end{equation}
where $\z$ is the vorticity, $\psi$ is the \iftoggle{geocoords}{streamfunction (electric potential)}{electric potential (streamfunction)}, and $\v{v} = \unit{z} \times \nabla \psi$ is the fluid velocity.
	\iftoggle{geocoords}{The conventional geophysical coordinate system is used where $\b$ is the local gradient of Coriolis parameter (plasma density) in the $y$ direction and $x$ is the zonal direction.  $f$ and $D$ represent forcing and dissipation.  Variables can be decomposed into mean and fluctuating components, e.g., $\z = \ol{\z} + \widetilde{\z}$, where $\ol{\z} = L_x^{-1} \int_0^{L_x} dx\, \z$ is a zonally averaged quantity.  The Generalized Hasegawa--Mima equation involves taking $\widetilde{\z} = \nablabarsq \widetilde{\psi} = (\nabla^2 - \LD)\widetilde{\psi}$ for fluctuations and $\ol{\z} = \nabla^2 \ol{\psi}$ for the mean flow, where $L_d$ is the deformation radius (plasma sound radius).  The barotropic vorticity equation is recovered for $\LD = 0$.}{The conventional plasma coordinate system is used where $\k$ is the local gradient of plasma density (Coriolis parameter) in the $-x$ direction and $y$ is the zonal direction.  $f$ and $D$ represent forcing and dissipation.  Variables can be decomposed into mean and fluctuating components, e.g., $\z = \ol{\z} + \widetilde{\z}$, where $\ol{\z} = L_y^{-1} \int_0^{L_y} dy\, \z$ is a zonally averaged quantity.  The Generalized Hasegawa--Mima equation involves taking $\widetilde{\z} = \nablabarsq \widetilde{\psi} = (\nabla^2 - \LD)\widetilde{\psi}$ for fluctuations and $\ol{\z} = \nabla^2 \ol{\psi}$ for the mean flow, where $\r_s$ is the plasma sound radius (deformation radius).  In plasma coordinates, lengths are normalized to make $\r_s=1$; the geophysical barotropic vorticity equation is recovered for $\LD = 0$.}

The quasilinear system is obtained by, within the equation for the fluctuations, discarding the terms quadratic in fluctuations.
	
We take the forcing to consist of statistically homogeneous white noise with covariance $F$ and dissipation to consist of a constant friction $\m$.  Viscosity is not necessary to regularize the dynamics in the quasilinear system because there is no turbulent cascade.  Direct numerical simulations of both the original system and the quasilinear approximation exhibit steady zonal flows \cite{srinivasan:2012,parker:2013}.

We quote the CE2 equations \cite{srinivasan:2012,parker:2013}:
	\begin{subequations}
	\begin{gather}
	\iftoggle{geocoords}{
		\partial_t W(x,y,\ybar,t) + (U_+ - U_-) \partial_x W - (U_+'' - U_-'') \left( \nablabarsq + \frac{1}{4} \partial_\ybar^2 \right) \partial_x \Psi \notag \\
			\qquad \qquad \qquad \qquad -[2\b - (U_+'' + U_-'')] \partial_\ybar \partial_y \partial_x \Psi = F(x,y) - 2\m W \label{CE2_W}, \\
			\partial_t U(\ybar, t) + \m U = -\partial_\ybar \partial_y \partial_x \Psi(x,y,\ybar, t) |_{(x,y)=(0,0)}, \label{CE2_U}
	}{
		\partial_t W(x,y,\xbar,t) + (U_+ - U_-) \partial_y W - (U_+'' - U_-'') \left( \nablabarsq + \frac{1}{4} \partial_\xbar^2 \right) \partial_y \Psi \notag \\
			\qquad \qquad \qquad \qquad +[2\k + (U_+'' + U_-'')] \partial_\xbar \partial_x \partial_y \Psi = F(x,y) - 2\m W \label{CE2_W}, \\
			\partial_t U(\xbar, t) + \m U = \partial_\xbar \partial_x \partial_y \Psi(x,y,\xbar, t) |_{(x,y)=(0,0)}, \label{CE2_U}
		}
	\end{gather}
	\end{subequations}
\iftoggle{geocoords}{where $U$ is the zonal flow velocity, $U_\pm = U(\ybar \pm y/2, t)$, and $U'' = \partial_\ybar^2 U$.  $W$ and $\Psi$ are the two-point, one-time covariance of vorticity and streamfunction, respectively, and are related by $W(x,y,\ybar,t) = L_+ L_- \Psi(x,y,\ybar,t)$, where $L_\pm = \nablabarsq \pm \partial_y \partial_\ybar + \tfrac{1}{4} \partial_\ybar^2$.}{where $U$ is the zonal flow velocity, $U_\pm = U(\xbar \pm x/2, t)$, and $U'' = \partial_\xbar^2 U$.  $W$ and $\Psi$ are the two-point, one-time covariance of vorticity and streamfunction, respectively, and are related by $W(x,y,\xbar,t) = L_+ L_- \Psi(x,y,\xbar,t)$, where $L_\pm = \nablabarsq \pm \partial_x \partial_\xbar + \tfrac{1}{4} \partial_\xbar^2$.}
Equivalently, $W$ is the physical-space Wigner function for $\z$ and \eqnref{CE2_W} is the Wigner transport equation that can be alternatively described in terms of formal Weyl symbols \cite{krommes:book}.  The right-hand-side of \eqnref{CE2_U} is the mean force due to the divergence of the Reynolds stress.
\iftoggle{geocoords}{Periodicity is assumed in $\ybar$, the inhomogeneity coordinate.}{Periodicity is assumed in $\xbar$, the inhomogeneity coordinate.}

\paragraph{Asymptotic wave kinetic equation}
A wave kinetic equation (WKE) takes the form
	\begin{equation}
		\partial_t \N(\v{k}, \v{x}, t) + \pd{\w}{\v{k}} \cdot \pd{\N}{\v{x}} - \pd{\w}{\v{x}} \cdot \pd{\N}{\v{k}} = S,
		\label{wke_general}
	\end{equation}
where $\N$ is the wave action density and $S$ represents sources and sinks \cite{krommes:book, connaughton:2015}.  With $S=0$, it is Hamiltonian with $\w$ generating the equations for wavepacket trajectories through phase space, $d\v{x}/dt = \nabla_\v{k} \w$ and $d\v{k}/dt = -\nabla_\v{x} \w$.
\iftoggle{geocoords}{For the physical space coordinates, we write $\v{x} = (\ol{x}, \ol{y}) \to \ybar$ because we only allow inhomogeneity in one direction.}{For the physical space coordinates, we write $\v{x} = (\ol{x}, \ol{y}) \to \xbar$ because we only allow inhomogeneity in one direction.}

A form which has been used as the starting point in many plasma physics studies assumes the zonal flows are \emph{asymptotically} large scale relative to the fluctuations, in which case
\iftoggle{geocoords}{$\w(\v{k}, \ybar) = -\b k_x/\kbsq + k_x U(\ybar)$}{$\w(\v{k}, \xbar) = \k k_y/\kbsq + k_y U(\xbar)$}
, where $\kbsq = k^2 + \LD$ and $k^2 = k_x^2 + k_y^2$.  Equation \eqref{wke_general} becomes 
	\begin{subequations}
		\label{asymptotic_wke}
	\begin{equation}
		\iftoggle{geocoords}{
		\partial_t \N(\v{k}, \ybar, t) - k_x U' \pd{\N}{k_y} + \frac{2 \b k_x k_y}{\ol{k}^4} \pd{\N}{\ybar} = \frac{F(\v{k})}{\b} - 2\m \N,
		}{
		\partial_t \N(\v{k}, \xbar, t) - k_x U' \pd{\N}{k_x} - \frac{2 \k k_x k_y}{\ol{k}^4} \pd{\N}{\xbar} = \frac{F(\v{k})}{\b} - 2\m \N,
		}
		\label{asymptotic_wke_W}
	\end{equation}
where forcing and dissipation have been inserted and the wave action density is \iftoggle{geocoords}{$\N = W/\b$}{$\N = W/\k$}.  Dynamics are closed by adding the equation for the zonal flow, 
	\begin{equation}
	\iftoggle{geocoords}{
		\partial_t U(\ybar, t) + \m U = \partial_\ybar \int \frac{d\v{k}}{(2\pi)^2} \frac{\b k_x k_y}{\ol{k}^4} \N(\v{k}, \ybar, t),
		}{
		\partial_t U(\xbar, t) + \m U = -\partial_\xbar \int \frac{d\v{k}}{(2\pi)^2} \frac{\k k_x k_y}{\ol{k}^4} \N(\v{k}, \xbar, t),}
		\label{asymptotic_wke_U}
	\end{equation}
	\end{subequations}
where we have used the Fourier transform convention $f(k) = \int dx\,  e^{-ikx} f(x)$.  We call this coupled set the Asymptotic WKE.  With no forcing or dissipation, it conserves enstrophy within the fluctuations only and conserves energy between the fluctuations and zonal flow.

\begin{figure}
		\centering
		\includegraphics{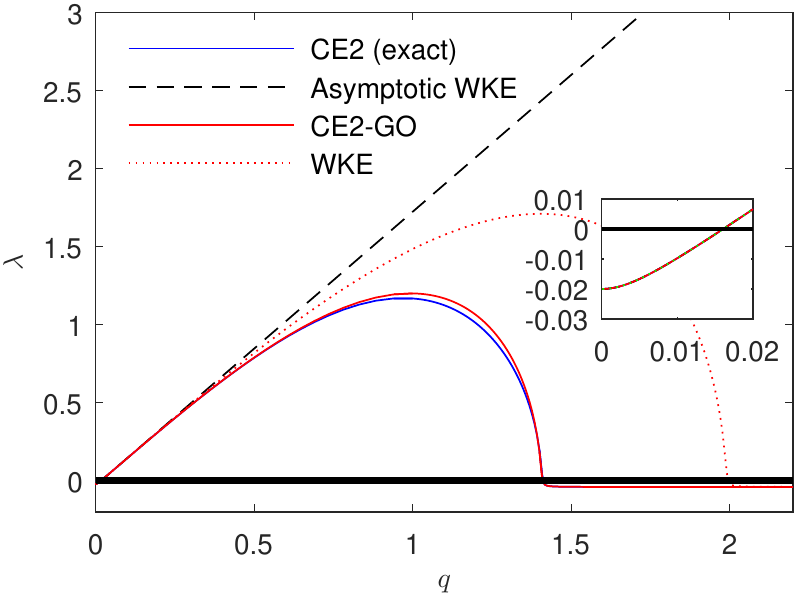}
		\caption{Dispersion relation of zonostrophic instability describing linear stage of growth of zonal flows about a homogeneous equilibrium.  Inset: zoomed in at small $q$, where all the curves overlap.   For the calculations of the dispersion relation, see the Supplemental Material.  Parameters: $\m=0.02$, $\b=1$, $\LD=1$, $F = 4\pi \ve k_f \de(k-k_f)$,  $\ve=1$, $k_f=1$.}
		\label{fig:disprelation}
	\end{figure}

The Asymptotic WKE is also pathological.  It drives growth of arbitrarily small-scale zonal flow, violating its fundamental assumption of scale separation as well as precluding any kind of well-behaved solution.  The dispersion relation of zonostrophic instability for perturbations \iftoggle{geocoords}{$\sim e^{\l t} e^{i q \ybar}$}{$\sim e^{\l t} e^{i q \xbar}$} is shown in \figref{fig:disprelation}.

The dispersion relation plotted in \figref{fig:disprelation} is calculated as follows.  \iftoggle{geocoords}{Equation \eref{asymptotic_wke} contains a homogeneous equilibrium, independent of $\ybar$, at $\N_H = F/2\b \m$, $U=0$.  One writes $\N = N_H + e^{iq\ybar} e^{\l t} \N_1(k_x, k_y)$, $U = e^{iq\ybar} e^{\l t} U_1$ and linearizes.}{Equation \eref{asymptotic_wke} contains a homogeneous equilibrium, independent of $\xbar$, at $\N_H = F/2\k \m$, $U=0$.  One writes $\N = N_H + e^{iq\xbar} e^{\l t} \N_1(k_x, k_y)$, $U = e^{iq\xbar} e^{\l t} U_1$ and linearizes.}
The linearized form of \eqnref{asymptotic_wke_W} can be solved for $\N_1$ in terms of $U_1$, and then substituted into \eqnref{asymptotic_wke_U}, yielding a single nonlinear equation for the eigenvalue $\l$:
	\begin{equation}
	\iftoggle{geocoords}{
		\l + \m = -q^2 \int \frac{d\v{k}}{(2\pi)^2} \frac{k_x^2 k_y}{(\l + 2\m)\ol{k}^4 + 2i \b q k_x k_y} \b \pd{\N_H}{k_y}
	}{
		\l + \m = -q^2 \int \frac{d\v{k}}{(2\pi)^2} \frac{k_x k_y^2}{(\l + 2\m)\ol{k}^4 - 2i \k q k_x k_y} \k \pd{\N_H}{k_x}
	}
		\label{disprelation_asymptotic_wke}
	\end{equation}
When $q$ is large, $\l \sim q$ and grows without bound.  For more details, see the Supplemental Material\footnote{Supplemental Material available online \href{http://nbviewer.jupyter.org/github/jeffbparker/ZonalFlowWaveKinetic/blob/master/Supplemental_Material.ipynb}{here}.}.

\begin{figure}
		\centering
		\includegraphics{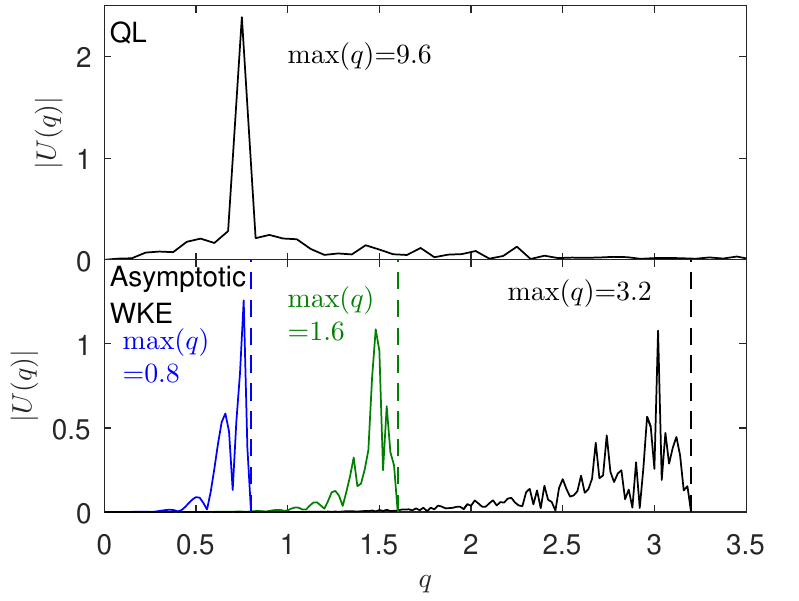}
		\caption{Spectrum of the zonal flow $U(q)$ [the Fourier transform of $U(\ybar)$] in the nonlinear saturated state at $t_f = 6/\m$ in simulations of the quasilinear system and the Asymptotic WKE.  In each simulation, the zonal flows reach a steady state.  For the asymptotic WKE, the results from three simulations are shown, with three different values for the maximum resolved wavenumber of the zonal flow.  In each case, the zonal flow energy concentrates at the highest resolved wavenumbers.  Same parameters as in \figref{fig:disprelation}.}
		\label{fig:U_kybar}
	\end{figure}

Nonlinear simulations of the Asymptotic WKE confirm its pathological behavior.  Figure \ref{fig:U_kybar} shows the spectrum of the zonal flow in the saturated state for different values of the highest resolved wavenumber $q$.  No matter the resolution, the highest modes grow fastest and zonal flow energy concentrates in the highest resolved wavenumbers.  For comparison, the figure also shows the same plot for a direct numerical simulation of the quasilinear system.  These simulations are pseudospectral, dealiased, and nonlinearly conserve energy and enstrophy to machine precision.
%The code is available online\footnote{The code will be made available upon publication.}.

For very small $q$, this model correctly predicts for zonostrophic instability that fluctuations provide an effective forcing proportional to $q^2$.  Figure \ref{fig:disprelation} shows that the $q^2$ regime only exists for extremely small $q$ and may be wholly irrelevant in practice because zonal flows are not so large scale. 

Some studies have been careful to not carry the Asymptotic WKE too far, concluding from the $q^2$ regime only that the symmetry-breaking instability is likely a generic phenomenon, which was an early insight \cite{lebedev:1995,smolyakov:2000b,krommes:2000}.  Others, however, have attempted to use the Asymptotic WKE to predict growth rates of zonal flow or to seek nonlinear solutions of zonal flow-turbulence interaction, and that is invalid from the start \cite{manin:1994,diamond:1998,smolyakov:2000,malkov:2001a,malkov:2001b,malkov:2001c,itoh:2004,itoh:2005b,anderson:2002,anderson:2006,wang:2009,kaw:2002,singh:2014,balescu:2003,trines:2005}.  It appears all but two of these papers \cite{manin:1994, trines:2005} do not discuss numerical solutions of either the Asymptotic WKE or of the dispersion relation \eqnref{disprelation_asymptotic_wke} but instead only deal with approximations.  One paper provides numerical simulations of the Asymptotic WKE and reports singularities due to the instability at large $q$, then applies a scheme to numerically suppress the singularities \cite{manin:1994}.

Approximations of the Asymptotic WKE are dangerous because they elide its pathology.  Sometimes, a random-walk argument based on random zonal flows has been invoked to transform \eqnref{asymptotic_wke_W} into a diffusion equation \cite{diamond:1998, malkov:2001b}.  However, numerical simulations of Hasegawa--Mima or Hasegawa--Wakatani models exhibit steady, not random, zonal flows \cite{parker:2013,numata:2007}.

% A quick note on explaining the WKE plot.  When the linearization of the term $F/(\b-U'')$ was included in Eq (), some very incorrect results were obtained, including the eigenvalues turning complex at some values of $q$.  What is shown is the result when the linearization of $F/(\b-U'')$ is ignored, which is justifiable in an alternative procedure.  
	
\paragraph{Geometrical optics approximations}
We report two reductions of CE2 involving a geometrical optics approximation.

The first approximation, which we call CE2-GO, involves an expansion in space that is justified when inhomogeneities are weak.  The $U_\pm$ terms in \eqnref{CE2_W} are Taylor expanded for 
\iftoggle{geocoords}{small $y$ and only lowest order in $\partial_\ybar$ is kept, e.g., the lowest order relation $W(x,y,\ybar) = \ol{\nabla}^4 \Psi (x,y,\ybar)$ is used.}{small $x$ and only lowest order in $\partial_\xbar$ is kept, e.g., the lowest order relation $W(x,y,\xbar) = \ol{\nabla}^4 \Psi (x,y,\xbar)$ is used.}
After Fourier transforming $(x,y) \to \v{k}$, we obtain
		\begin{subequations}
		\begin{gather}
		\iftoggle{geocoords}{
			\partial_t W(\v{k},\ybar,t) - k_x U'\pd{W}{k_y} - k_x U''' \pd{}{k_y} \left( \frac{W}{\kbsq} \right) + 2(\b - U'') \frac{k_x k_y}{\ol{k}^4} \pd{W}{\ybar} = F(\v{k}) - 2\m W, \label{CE2-GO_W}\\
			\partial_t U(\ybar, t) + \m U = \partial_\ybar \int \frac{d\v{k}}{(2\pi)^2} \frac{k_x k_y}{\ol{k}^4} W.
		}{
			\partial_t W(\v{k},\xbar,t) - k_y U'\pd{W}{k_x} - k_y U''' \pd{}{k_x} \left( \frac{W}{\kbsq} \right) - 2(\k + U'') \frac{k_x k_y}{\ol{k}^4} \pd{W}{\xbar} = F(\v{k}) - 2\m W, \label{CE2-GO_W}\\
			\partial_t U(\xbar, t) + \m U = -\partial_\xbar \int \frac{d\v{k}}{(2\pi)^2} \frac{k_x k_y}{\ol{k}^4} W.
		}
		\end{gather}
		\end{subequations}
	
In the second approximation, the zonal flow is assumed quasi-static, leading to conservation of wave action density \iftoggle{geocoords}{$\N = W/(\b - U'')$.  A wave kinetic equation in the form of \eqnref{wke_general} is derived from \eqnref{CE2-GO_W} by dividing by $(\b - U'')$.}{$\N = W/(\k + U'')$.  A wave kinetic equation in the form of \eqnref{wke_general} is derived from \eqnref{CE2-GO_W} by dividing by $(\k + U'')$.}
One obtains
	\begin{subequations}
		\begin{gather}
		\iftoggle{geocoords}{
			\partial_t \N(\v{k},\ybar,t) - k_x \left( U' + \frac{U'''}{\kbsq} \right) \pd{\N}{k_y} + 2(\b - U'') \frac{k_x k_y}{\ol{k}^4} \pd{\N}{\ybar} = \frac{F(\v{k})}{\b - U''} - 2\m \N, \label{WKE_new_N} \\
			\partial_t U(\ybar, t) + \m U = \partial_\ybar \int \frac{d\v{k}}{(2\pi)^2} \frac{k_x k_y}{\ol{k}^4} (\b - U'') \N.
		}{
				\partial_t \N(\v{k},\xbar,t) - k_y \left( U' + \frac{U'''}{\kbsq} \right) \pd{\N}{k_x} - 2(\k + U'') \frac{k_x k_y}{\ol{k}^4} \pd{\N}{\xbar} = \frac{F(\v{k})}{\k + U''} - 2\m \N, \label{WKE_new_N} \\
			\partial_t U(\xbar, t) + \m U = -\partial_\xbar \int \frac{d\v{k}}{(2\pi)^2} \frac{k_x k_y}{\ol{k}^4} (\k + U'') \N.
		}
		\end{gather}
		\end{subequations}
We denote this coupled set as the WKE \cite{wordsworth:2009}.  Equation \eref{WKE_new_N} arises from a wave frequency \iftoggle{geocoords}{$\w(\v{k},\ybar) = -k_x (\b - U''(\ybar))/\kbsq + k_x U(\ybar)$}{$\w(\v{k},\xbar) = k_y (\k + U''(\ybar))/\kbsq + k_y U(\ybar)$}
, which follows directly from the quasilinear equations; the $U''$ term comes from $\v{\widetilde{v}} \cdot \nabla \ol{\z}$.  This WKE is invalid when $U''$ is rapidly evolving such as during an instability because wave action is not conserved.  A wave kinetic formulation therefore may not be used to study the dynamics in which a zonal flow evolves to a steady state.  However, in the quasi-steady-state limit, the WKE is valid and equivalent to CE2-GO.  The WKE exhibits a close connection to the Rayleigh--Kuo stability criterion, which states that a necessary condition for instability of fixed mean flow is that \iftoggle{geocoords}{$\b - U''$ has a zero somewhere \cite{kuo:1949}; such a condition would create singularities in $\N = W/(\b-U'')$.}{$\k + U''$ has a zero somewhere \cite{kuo:1949}; such a condition would create singularities in $\N = W/(\k+U'')$.}

Both CE2-GO and the WKE nonlinearly conserve enstrophy within the fluctuations and conserve energy between the fluctuations and zonal flow, although for the WKE one again must assume quasi-static zonal flow.  Within the geometrical optics approximation, the energy of the fluctuations is \iftoggle{geocoords}{$(2L_y)^{-1} \int_0^{L_y} d\ybar \int d\v{k}\, (2\pi)^{-2} W(\v{k}, \ybar) / \kbsq$, and the enstrophy is the same integral with $W$ instead of $W / \kbsq$.  The energy of the zonal flow is $(2L_y)^{-1} \int_0^{L_y} d\ybar\, U(\ybar)^2$.}{$(2L_x)^{-1} \int_0^{L_x} d\xbar \int d\v{k}\, (2\pi)^{-2} W(\v{k}, \xbar) / \kbsq$, and the enstrophy is the same integral with $W$ instead of $W / \kbsq$.  The energy of the zonal flow is $(2L_x)^{-1} \int_0^{L_x} d\ybar\, U(\xbar)^2$.}

In  the limit of asymptotically large-scale zonal flows in which $U''$ and $U'''$ are neglected, both CE2-GO and the WKE reduce to the Asymptotic WKE.

The dispersion relations for zonostrophic instability in CE2-GO and the WKE are shown in \figref{fig:disprelation}.  CE2-GO agrees remarkably well with the exact CE2, even at $q$ where a scale separation is not satisfied.  The $U'''$ term in \eqnref{CE2-GO_W} and \eqnref{WKE_new_N} is essential; when only the flow shear $U'$ is retained, the dispersion relation grows linearly in $q$.  The WKE does not correctly predict the dispersion relation because its fundamental assumption of quasi-static zonal flow is violated.

We emphasize that CE2 is exact for quasilinear dynamics while CE2-GO is approximate, so CE2--GO should not be preferred just because of its simpler and more intuitive form.  Numerical solutions of CE2 are feasible and many studies have been carried out successfully \cite{squire:2015,farrell:2003,farrell:2007,constantinou:2014,marston:2008,tobias:2011,parker:2013,parker:2014}.  Moreover, numerically solving CE2 as compared to CE2--GO takes comparable effort as both are three dimensional, although perhaps the geometrical optics approximation reduces the computational cost.  Certain analytic calculations are also possible for CE2 \cite{srinivasan:2012,bakas:2013b,bakas:2015,parker:2013,parker:2014}.
	
The utility of CE2-GO and the WKE, if they prove to be accurate, will come from their transparent phase-space formulation, more intuitive interpretation of wave dynamics, and greater possibility of analytic manipulation as compared to CE2.  The WKE is valid for studying steady states, and we anticipate it will provide new insight into how wavepackets behave in a steady mean flow.

\paragraph{Conclusions}
The wave kinetic equation where zonal flows are assumed to be asymptotically large scale compared to the fluctuations is mathematically pathological.  Within this approximation, the instability that drives zonal flows grows linearly in wavenumber without bound.  Nonlinear numerical solutions confirm that zonal flow energy concentrates in the highest resolved wavenumbers.  For well-behaved equations, the effect of mean flow on the fluctuations cannot be described solely in terms of the local mean flow shear.

CE2 is an exact description of quasilinear dynamics, and we have described two geometrical optics approximations that offer simpler and more intuitive phase-space equations.  These formulations involve not only the local mean flow shear, but also the second and third derivative of the mean flow.  The first formulation, CE2-GO, requires only a spatial scale expansion and agrees with the exact zonostrophic instability dispersion relation remarkably well.  The second formulation, a wave kinetic equation written in terms of conservation of wave action, is valid only when the zonal flow is slowly varying.  This requirement precludes the use of the wave kinetic equation for studying the important dynamics that determine how zonal flows grow and saturate, during which wave action is not conserved.

Future work will continue to study the geometrical optics approximations and their validity.

\begin{acknowledgments}
Useful discussions with Navid Constantinou, Ilya Dodin, and John Krommes are acknowledged.
\end{acknowledgments}

\bibliographystyle{apsrev4-1}
\bibliography{failurewke} % if running BiBTeX

\end{document}